\documentclass[reprint,aps,pra,superscriptaddress]{revtex4-2}
\usepackage[utf8]{inputenc}

\usepackage{xcolor}

\definecolor{algogreen}{RGB}{46, 106, 107}

\usepackage[colorlinks=true,
bookmarks=false,
linkcolor=algogreen,
urlcolor=algogreen,
citecolor=algogreen,
breaklinks]{hyperref}

\usepackage{graphicx}

\usepackage{latexsym,amsmath}
\usepackage{amsbsy}
\usepackage{float}
\usepackage{hyperref}
\usepackage[normalem]{ulem}
\usepackage{natbib}
\usepackage{todonotes}
\usepackage{comment}

\usepackage[utf8]{inputenc}
\usepackage[T1]{fontenc}
\usepackage[sc,osf]{mathpazo}
\usepackage{graphicx}
\usepackage{amssymb}
\usepackage{color}
\usepackage{amsthm}
\usepackage{psfrag}
\usepackage{ifsym}
\usepackage{epstopdf}
\usepackage{dsfont}
\usepackage{amsfonts}
\usepackage{multirow}
\usepackage{physics}
\usepackage{bm}
\usepackage{bbm}
\usepackage{mathtools}

\def\1{\mathbf{1}}
\def\0{\mathbf{0}}

\newcommand{\eg}{{\it{e.g.~}}}
\newcommand{\ie}{{\it{i.e.~}}}

\newcommand{\beq}{\begin{equation}}
\newcommand{\eeq}{\end{equation}}
\newcommand{\bea}[1]{\begin{equation}\begin{array}{#1}}
\newcommand{\eea}{\end{array}\end{equation}}
\newcommand{\beqn}{\begin{eqnarray}}
\newcommand{\eeqn}{\end{eqnarray}}

\renewcommand{\rho}{\varrho}

\newcommand{\processnext}[1]{%
  \ifx\listfinish#1\empty\else\listact{#1}\expandafter\processnext\fi}

\newcommand{\ea}{\end{eqnarray}}
\newcommand{\ban}{\begin{eqnarray*}}
\newcommand{\ean}{\end{eqnarray*}}

\begin{document}

\title{Virtual linear map algorithm\\ for classical boost in near-term quantum computing}

\author{Guillermo Garc\'{i}a-P\'{e}rez}
\email{guille@algorithmiq.fi}
\affiliation{Algorithmiq Ltd, Kanavakatu 3 C, FI-00160 Helsinki, Finland}

\author{Elsi-Mari Borrelli}
\affiliation{Algorithmiq Ltd, Kanavakatu 3 C, FI-00160 Helsinki, Finland}

\author{Matea Leahy}
\affiliation{Algorithmiq Ltd, Kanavakatu 3 C, FI-00160 Helsinki, Finland}

\author{Joonas Malmi}
\affiliation{Algorithmiq Ltd, Kanavakatu 3 C, FI-00160 Helsinki, Finland}

\author{Sabrina Maniscalco}
\affiliation{Algorithmiq Ltd, Kanavakatu 3 C, FI-00160 Helsinki, Finland}

\author{Matteo A. C. Rossi}
\affiliation{Algorithmiq Ltd, Kanavakatu 3 C, FI-00160 Helsinki, Finland}

\author{Boris Sokolov}
\affiliation{Algorithmiq Ltd, Kanavakatu 3 C, FI-00160 Helsinki, Finland}

\author{Daniel Cavalcanti}
\affiliation{Algorithmiq Ltd, Kanavakatu 3 C, FI-00160 Helsinki, Finland}

\date{\today}

\begin{abstract}
The rapid progress in quantum computing witnessed in recent years has sparked widespread interest in developing scalable quantum information theoretic methods to work with large quantum systems.
For instance, several approaches have been proposed to bypass tomographic state reconstruction, and yet retain to a certain extent the capability to estimate multiple physical properties of a given state previously measured.
In this paper, we introduce the Virtual Linear Map Algorithm (VILMA), a new method that enables not only to estimate multiple operator averages using classical post-processing of informationally complete measurement outcomes, but also to do so for the image of the measured reference state under low-depth circuits of arbitrary, not necessarily physical, $k$-local maps.
We also show that VILMA allows for the variational optimisation of the virtual circuit through sequences of efficient linear programs.
Finally, we explore the purely classical version of the algorithm, in which the input state is a state with a classically  efficient representation, and show that the method can prepare ground states of many-body Hamiltonians.
\end{abstract}

\maketitle

\section{Introduction}\label{sec:intro}
The last decade has witnessed a tremendous progress in the field of quantum computing. Today, we have access to several devices composed by tens to hundreds of physical qubits that can be used as test-beds for small quantum simulations and even to demonstrate the superiority of quantum information processing over its classical counterpart \cite{Google_2019,Zhong_2020,Madsen2022}. However, the small number of qubits, limited connectivity, and presence of noise remain strong barriers for the implementation of quantum protocols offering true practical advantage over classical computing. Due to the limitations in the quantum hardware, hybrid algorithms incorporating classical pre- or post-processing methodologies are often utilised~\cite{endo_hybrid_2021,bharti_noisy_2022}.
Despite the promise that such techniques hold in demonstrating the first useful applications for quantum computing in the near future, so far such showcases are still missing. Thus, the development of further classical pre- and post-processing techniques remains of central importance for the future of quantum computing and simulation~\cite{urbanek2021mitigating,wallman2016noise,li_efficient_2017,temme_error_2017,endo_practical_2018,smart_quantum-classical_2019,kandala_error_2019,Huang_2020,suchsland_algorithmic_2021,Aolita_2022,ravi_cafqa_2022,rakyta_efficient_2022}.

A key limitation that we encounter when post-processing the result of a quantum computation is the poor statistics obtained in such calculations. Due to the exponential growth of the Hilbert space, the number of shots one can access in a typical experiment is not enough to perform quantum state tomography. This prevents us from studying many properties of the quantum state produced by the quantum processor. Even determining the expected value of relevant observables within a satisfactory accuracy may be difficult in this regime~\cite{mcclean_exploiting_2014,wecker_progress_2015,babbush_low-depth_2018,cai_resource_2020}. Because of this, alternative estimation techniques must be developed~\cite{Cramer_2010,da_Silva_2011,torlai_neural-network_2018,Carrasquilla_2019,Paini_2019,morris_selective_2020,Huang_2020,Jiang_2020,huggins_efficient_2021,Garc_a_P_rez_2021,Morris_2022}

In this article we introduce a classical algorithm that can be used as a post-processing method to assist computation on quantum devices. The main goal of the algorithm is to estimate physical properties of a modified state corresponding to the result of a virtual transformation applied on the state produced by the quantum processor. By virtual, we mean that these operations are not physically implemented in the quantum system, but rather implemented in a classical device. As such, these virtual maps can generally represent non-physical operations, that is, they do not need to be described by completely positive maps. The key feature of the algorithm is the fact that its input is the statistics obtained from an informationally-complete set of measurements applied on a quantum state, but does not require full state tomography.  We call this algorithm VILMA, from \emph{Virtual Linear Map Algorithm}.

The paper is structured as follows: In Section \ref{sec:method} we give a general overview of the algorithm, formalise the VILMA method and provide some mathematical details.
In Section \ref{sec:numerics} we show some numerical results on the reconstruction of multiple expectation values on transformed states.
Section \ref{sec:optimisation} presents a methodology to optimise the VILMA maps. 
In Section \ref{sec:classicalVILMA} we discuss VILMA as a purely classical method and, finally, in Section \ref{sec:conclusion} we present our conclusion.
\section{The VILMA method}\label{sec:method}

\begin{figure}
    \centering
    \includegraphics[width= \columnwidth]{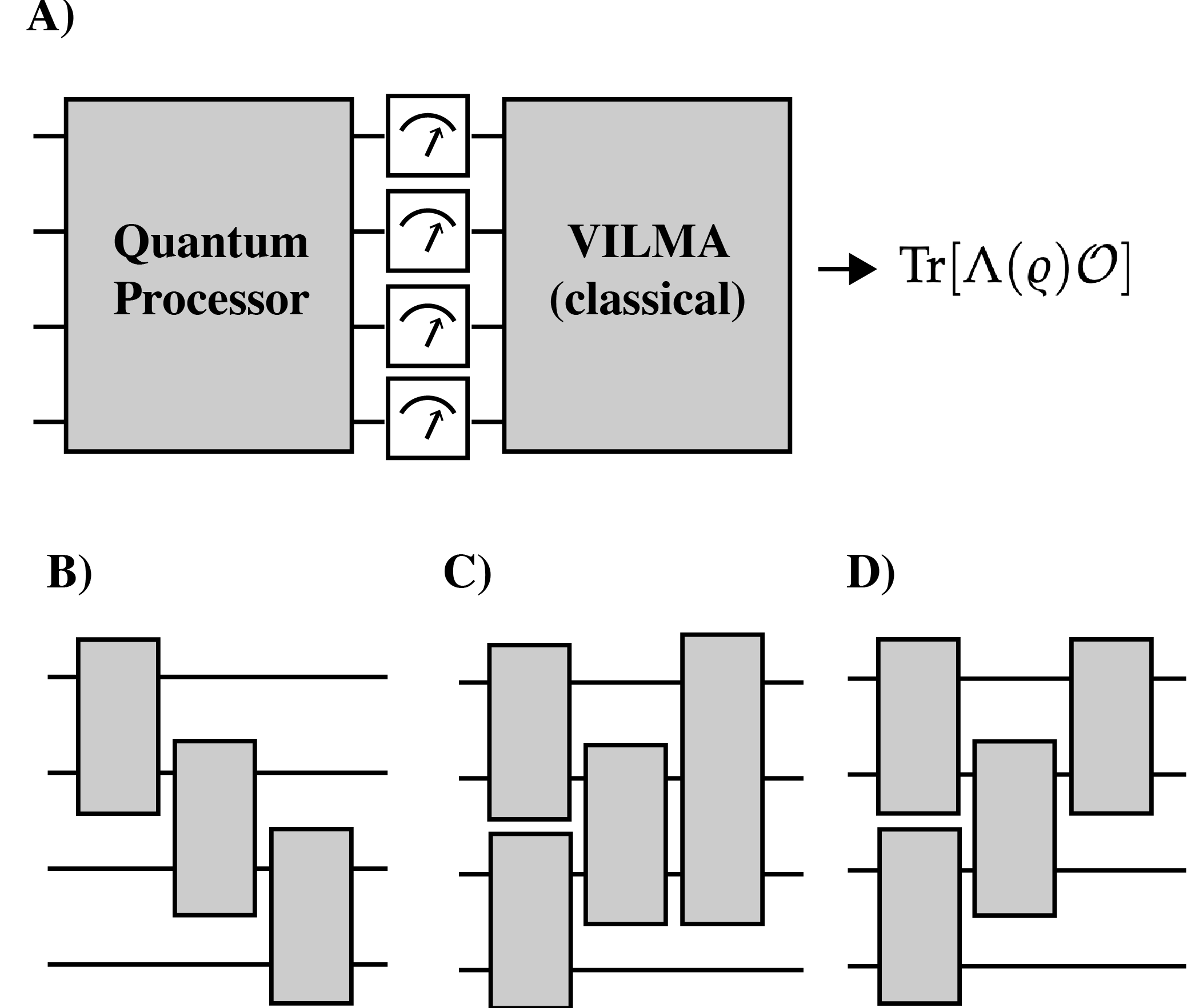}
    \caption{A) A quantum processor produces an $N$-qubit state $\rho$ that is then measured by a tomographically complete set of measurements. The measurement statistics is then post-processed classically through VILMA to provide an estimate of $\mathrm{Tr} [\Lambda(\rho) \mathcal{O}]$, where $\Lambda$ is a linear map and $\mathcal{O}$ an efficiently represented observable. B-D) Examples of lower depth circuits composed by a sequence of linear maps acting on few qubits that can be used to define VILMA.}
    \label{fig:maps}
\end{figure}

Suppose we have at our disposal a quantum processor that can produce an $N$-qubit state $\rho$. To characterise the system fully, estimating the state $\rho$ is necessary, but becomes practically impossible even for tens of qubits. However, even if full state reconstruction is not possible one can still estimate the expected value $\mathrm{Tr} [\rho \mathcal{O}_i]$ of a set of observables $\{\mathcal{O}_i\}$. This can be done directly, \ie by measuring the observables, if these observables can be efficiently implemented in the physical apparatus. Or they can be indirectly estimated from the results of other measurements \cite{morris_selective_2020,Jiang_2020,Huang_2020,acharya_informationally_2021,Garc_a_P_rez_2021}, in the case that these observables have an efficient classical representation (for instance, $\mathcal{O}$ can be written as a linear combination of few Pauli strings). 

Here we go a step beyond and provide a method to estimate the expected value of measurements applied to a modified state $\rho'=\Lambda(\rho)$, where $\Lambda$ 
is a linear map. More specifically, VILMA allows us to estimate $\mathrm{Tr} [\Lambda(\rho) \mathcal{O}_i]$ given that we have the statistics of a tomographically complete set of measurements applied on $\rho$. We insist that due to the system's size we can only implement a low number of measurement rounds so we do not have enough statistics nor classical memory to reconstruct the state $\rho$. We also stress that in order to perform the computation efficiently on a classical computer, we need to restrict ourselves to observables that have an efficient classical representation, for instance, as a linear combination of few Pauli strings. This is the case for many problems of interest, such as estimating the ground state of local Hamiltonians.

We will also consider $\Lambda$ to be the composition of $k$-local maps forming a low-depth circuit. Examples of such maps can be seen in Fig.~\ref{fig:maps}B-D). As we will see, this assumption will allow us to not only carry on the calculations efficiently, but also to consider optimisation problems that variationally adjust each of these $k$-local maps. 

In what follows, we describe the methodology used in order to apply virtual post-measurement maps to the measured state.
It should be stated that, while the main goal of the method is to calculate quantities of the form $\mathrm{Tr}[\Lambda(\rho) \mathcal{O}]$ efficiently, the techniques can be easily generalised for other purposes.
We introduce some of these additional features as well.

\subsection{Operator averages on transformed states}

\subsubsection{Statistical estimation with IC measurement data}

Suppose the quantum processing unit (QPU) is in some $N$-qubit state $\rho$, which is measured by a $k$-local informationally complete (IC) measurement, that is given by an IC-POVM whose statistics singles out the state on which the measurement is applied~\cite{ariano_informationally_2004}.
By $k$-local here we mean that it is composed of POVM elements  that act non-trivially on $k$ qubits at most.
In principle, the IC-POVM need not be qubit-local nor minimal (\ie contain the minimum number of POVM elements), so that they can in principle act on $k > 1$ qubits and be overcomplete.
However, the number of qubits $k$ they act upon must be small enough so that operators of dimension $2^k \times 2^k$ can be dealt with classically.
Moreover, the number of POVM elements (which correspond to the number of outcomes of the measurement) should also be small enough so that one can store the measurement statistics in classical memory. In what follows we will assume $k = 1$ and that the POVMs have the minimal number of elements to be considered IC, although all the results included here can be easily generalised to larger $k$ and non-minimal measurements.
In particular, we will consider POVMs of the form $\{ \Pi_\mathbf{m} = \bigotimes_{i = 1}^N \Pi_{m_i}^{(i)} \}$, where each $\{ \Pi_{m_i}^{(i)} \}$ is a four-outcome IC-POVM on the Hilbert space of qubit $i$.

Given such a local IC-POVM, one can find a set of \textit{dual effects} $\{ D_{n_i}^{(i)} \}$ satisfying $\mathcal{O} = \sum_\mathbf{m} \mathrm{Tr}[\mathcal{O} \Pi_\mathbf{m}] D_\mathbf{m}$ for all $\mathcal{O}$.
A general method to compute the set of duals for a given POVM can be found in Ref.~\cite{guerini_quasiprobabilistic_2021}.
The decomposition of an $N$-qubit state $\rho$ in terms of these dual effects therefore reads $\rho = \sum_\mathbf{m} p_\mathbf{m} D_\mathbf{m}$, where $D_\mathbf{m} = \bigotimes_{i = 1}^N D_{m_i}^{(i)}$ and $p_\mathbf{m} = \mathrm{Tr}[\rho \Pi_\mathbf{m}]$ is the probability of obtaining outcome $\mathbf{m}$ once the measurement is performed.

We also assume that the observables of interest admit an efficient classical representations known to us.
In particular, we assume that they are of the form $\mathcal{O} = \sum_{\mathbf{k}} c_\mathbf{k} P_\mathbf{k}$, where $c_\mathbf{k} \in \mathbb{C}$ and $P_\mathbf{k} = \bigotimes_{i = 1}^{N} P_{k_i}^{(i)}$.
Each $P_{k_i}^{(i)}$ is a single-qubit operator, such as a Pauli operator.
In fact, in most applications, $\mathcal{O}$ is given in terms of such a linear combination of Pauli strings.
Importantly, for a wide class of relevant problems in many-body physics, the observables of interest can be mapped into low-weight $N$-qubit operators, meaning that only a small fraction of the Pauli operators in each string is different from identity.
Their corresponding observable averages can be estimated using local IC-POVM with polynomially scaling number of measurements~ \cite{Huang_2020,Jiang_2020,Garc_a_P_rez_2021}.

In experiments, only a finite number of measurement rounds $S$ can be performed. In this case such a procedure will result in a random sequence of outcomes $(\mathbf{m}_1, \ldots, \mathbf{m}_S)$, according to which we can write a crude approximation to the state, $\rho_S = \sum_{i = 1}^S D_{\mathbf{m}_i} / S$. Notice that $\lim_{S \rightarrow \infty} \rho_S = \rho$. 

Assuming a linear map $\Lambda$ that is independent of $\rho$ and  $\mathcal{O}$, 
we can write
\beqn
\mathrm{Tr}[\Lambda(\rho) \mathcal{O}] &=& \lim_{S \rightarrow \infty} \mathrm{Tr}[\Lambda(\rho_S) \mathcal{O}]\\
&=& \lim_{S \rightarrow \infty} \sum_{i = 1}^S \frac{1}{S} \sum_{\mathbf{k}} c_\mathbf{k} \mathrm{Tr}[\Lambda(D_{\mathbf{m}_i}) P_\mathbf{k}].
\eeqn
This implies that the quantity $\bar{\mathcal{O}}_{\Lambda} = \sum_{i = 1}^S \omega_{\mathbf{m}_i} / S$, with $\omega_{\mathbf{m}_i} = \sum_{\mathbf{k}} c_\mathbf{k} \mathrm{Tr}[\Lambda(D_{\mathbf{m}_i}) P_\mathbf{k}]$, is a consistent estimator of the mean value of $\mathcal{O}$ for state $\Lambda(\rho)$, that is, $\lim_{S \rightarrow \infty} \bar{\mathcal{O}}_{\Lambda} = \mathcal{O}_{\Lambda} \equiv \mathrm{Tr}[\Lambda(\rho) \mathcal{O}]$.
In addition, it is easy to see from the linearity of the expression that $\bar{\mathcal{O}}_{\Lambda}$ is also unbiased, meaning that its average over $S$-measurement sampling experiments is also equal to the observable average, $\mathbb{E}_S[\bar{\mathcal{O}}_{\Lambda}] = \mathcal{O}_{\Lambda}$.
Accordingly, the mean squared error of the estimation is given by $\mathbb{E}_S[(\bar{\mathcal{O}}_{\Lambda} - \mathcal{O}_{\Lambda})^2] = \mathrm{Var}(\omega_{\mathbf{m}}) / S$, where $\mathrm{Var}(\omega_{\mathbf{m}})$ is the variance of $\omega_{\mathbf{m}}$ over the probability distribution of the outcomes.
In realistic scenarios, the variance $\mathrm{Var}(\omega_{\mathbf{m}})$ is not known, but it can be estimated from the measurement outputs using the unbiased estimator $\bar{V}(\mathcal{O}_{\Lambda}) \equiv [\sum_{i = 1}^S \omega_{\mathbf{m}_i}^2 / S - (\bar{\mathcal{O}}_{\Lambda})^2] S / (S - 1)$.
In short, given $S$ IC-POVM outcomes from state $\rho$, computing $\bar{\mathcal{O}}_{\Lambda}$ and $\sigma \equiv \sqrt{\bar{V}(\mathcal{O}_{\Lambda}) / S}$ enables the estimation of the observable average $\mathcal{O}_{\Lambda}$ and the corresponding statistical error, respectively.
Regarding the latter, notice that while it is possible to derive analytical bounds for the statistical error based on the weight of the Pauli strings in the operator $\mathcal{O}$ when using IC POVM-based estimators \cite{Huang_2020,Jiang_2020,acharya_informationally_2021,Garc_a_P_rez_2021}, the calculations cannot be straightforwardly generalised to VILMA (in fact, the statistical error must depend on the details of the map $\Lambda$ as well).
It is nevertheless possible to guarantee the polynomial scaling of the variance for $k$-local observables and certain maps.

In practice, a major difficulty in dealing with the above terms lies in computing the terms $\mathrm{Tr}[\Lambda(D_{\mathbf{m}_i}) P_\mathbf{k}]$.
In some cases, this may be easy to do, for instance if the map $\Lambda(\cdot) = \sum_{ij} \lambda_{ij} B_i \cdot B_j^{\dagger}$ only involves a moderate number of terms $\lambda_{ij}$ and the operators $B_i$ have bounded locality; one can then simply compute $\mathrm{Tr}[D_{\mathbf{m}_i} \Lambda^{\dagger} (P_\mathbf{k})]$ in polynomial time, and the variance will be polynomially bounded too.
However, if the map has a complex structure, the computation may be much more challenging. In particular, notice that the explicit representation of $\Lambda(D_{\mathbf{m}_i})$ can generally be classically prohibitive, even for modest $N$.
In the next section, we provide an algorithm to compute these traces for a class of maps of particular relevance for quantum computing, namely circuits of $K$-local maps.
We show that, by restricting the circuit structure, we can make such a computation classically amenable.

\subsubsection{Traces involving mapped dual effects}

The main idea of VILMA is to restrict the structure and complexity of the map $\Lambda$ so that the calculation of each term $\mathrm{Tr}[\Lambda(D_{\mathbf{m}_i}) P_\mathbf{k}]$ only involves dealing with operators of bounded and efficient dimension during all the intermediate steps. More precisely, we ensure that the trace in $\mathrm{Tr}[\Lambda(D_{\mathbf{m}_i}) P_\mathbf{k}]$ can be computed through intermediate partial traces that keep the dimension of non-trivial operators (meaning, those that are not tensor products of single-qubit ones) under control. The details will become clear in what follows.

\begin{figure}[t]
    \centering
    \includegraphics[width=\columnwidth]{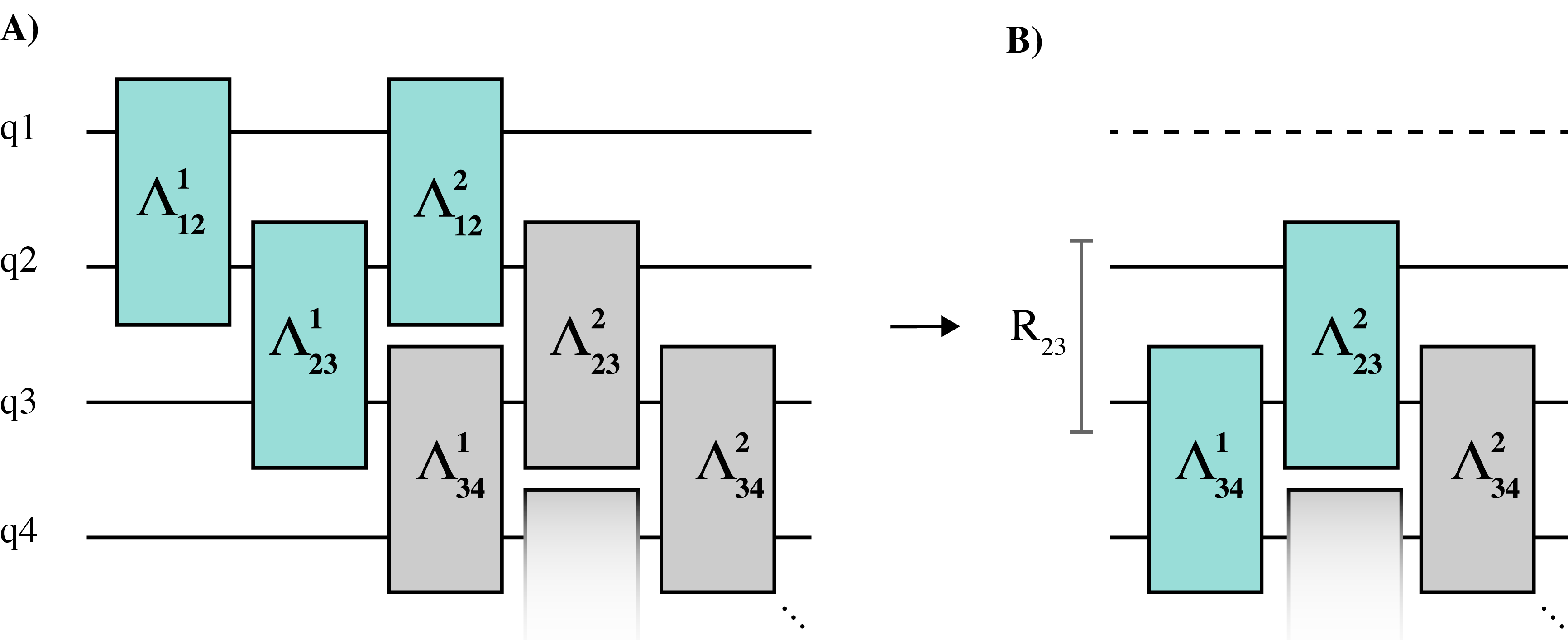}
    \caption{A circuit decomposition of a VILMA map composed by two layers of sequential 2-qubit gates. $q_k$ refers to qubit $k$. A) In the first step of the algorithm we apply the maps $\Lambda_{12}^1$, $\Lambda_{23}^2$ and $\Lambda_{12}^2$ (green boxes), which correspond to the causal cone of qubit 1 (see main text), to $D_{m_1} \otimes D_{m_2} \otimes D_{m_3}$. After multiplying with $P_{k_1} \otimes \mathbb{1} \otimes \mathbb{1}$ and tracing out qubit 1, we are left with a residual operator $R_{23}$ on qubits $2$ and $3$, and move to the next step. B) In the second step we apply maps $\Lambda_{34}^1$ and $\Lambda_{23}^2$, which correspond to the causal cone of qubit 2, to $R_{23} \otimes D_{m_4}$. 
   After multiplying the resulting operator with $P_{k_2} \otimes \mathbb{1} \otimes \mathbb{1}$, we trace out qubit 2 to obtain a residual operator $R_{34}$ on qubits $3$ and $4$, and move to the next step. The algorithm then proceeds similarly, by applying two-qubit maps on three-qubits and performing partial traces until all the maps are  applied and the traces are performed.
    }
    \label{fig:layers}
\end{figure}

We proceed by considering maps $\Lambda$ that can be decomposed in terms of a circuit of $k$-qubit maps with $k \leq K$ for some $K$, and with an adequate causal structure. For the sake of clarity, in what follows we will present this idea using a concrete example of maps composed of two layers of 2-qubit gates, shown in Figure \ref{fig:layers}:
\beq
\Lambda = \cdots \circ \Lambda_{34}^2 \circ \Lambda_{23}^2 \circ \Lambda_{12}^2 \circ \cdots \circ \Lambda_{34}^1 \circ \Lambda_{23}^1 \circ \Lambda_{12}^1.
\eeq
Indices $ij$ in $\Lambda_{ij}^l$ indicate that the map is applied to qubits $ij$, and index $l$ indicates that it is part of the $l^{th}$ layer. 
For such VILMA map, $\mathrm{Tr}[\Lambda(D_{\mathbf{m}_i}) P_\mathbf{k}]$ can be computed with no need for calculating any $m$-qubit operator with $m > 3$ at any point (even if the same structure is extended to $N>4$ qubits). To see how this can be done, first notice that the 2-qubit maps $\Lambda_{ij}^l$ acting on different qubits commute, so that we can write, for instance,
\beq
\Lambda = \cdots \circ  \Lambda_{23}^2 \circ \Lambda_{34}^1 \circ \Lambda_{12}^2 \circ \Lambda_{23}^1 \circ \Lambda_{12}^1. 
\eeq
This ordering presents the following advantage: the first three maps act non-trivially only on qubits 1, 2, and 3, and the rest of them act trivially on qubit 1.
Thus, we can first compute $\Lambda_{12}^2 \circ \Lambda_{23}^1 \circ \Lambda_{12}^1(D_{m_1} \otimes D_{m_2} \otimes D_{m_3})$, which requires the explicit computation of a three-qubit operator.
Now, the only non-trivial operation on qubit 1 is the multiplication with $P_{k_1}$, after which no other operation takes place in such Hilbert space when computing $\Lambda(D_{\mathbf{m}_i}) P_\mathbf{k}$.
Therefore, we can simply trace out qubit 1, and keep track of the resulting \textit{residual operator} 
\beq\label{eq:R23}
R_{2, 3} = \mathrm{Tr}_1 [\Lambda_{12}^2 \circ \Lambda_{23}^1 \circ \Lambda_{12}^1(D_{m_1} \otimes D_{m_2} \otimes D_{m_3}) P_{k_1} \otimes \mathbb{I}_{2} \otimes \mathbb{I}_{3}]
\eeq
In Appendix \ref{app:formalisation}, we present a more formal explanation of this step.

We can then proceed with maps $\Lambda_{34}^1$ and $\Lambda_{23}^2$, which must be applied to $R_{2, 3} \otimes D_{m_4}$, yielding another three-qubit operator.
Once again, we can trace out qubit 2 after multiplying the corresponding $P_{k_2}$ and keep track of the residual operator $R_{3, 4} = \mathrm{Tr}_2 [\Lambda_{23}^2 \circ \Lambda_{34}^1 (R_{2, 3} \otimes D_{m_4}) P_{k_2} \otimes \mathbb{I}_{3} \otimes \mathbb{I}_{4}]$.

This method can then be iterated until all the traces are performed.
Notice that the algorithm can be applied \textit{backwards} if we swap the roles of $D_{\mathbf{m}_i}$ and $P_{\mathbf{k}}$ and use the adjoint map $\Lambda^{\dagger}$, since $\mathrm{Tr} [\Lambda(D_{\mathbf{m}_i}) P_{\mathbf{k}}] = \mathrm{Tr} [\Lambda^\dagger(P_{\mathbf{k}}) D_{\mathbf{m}_i}]$.
The adjoint map $\Lambda^{\dagger}$ can be easily obtained by mirroring the circuit (\ie reversing the order of the maps) and substituting each map with its adjoint.
Depending on the map, the forwards or backwards algorithms can have different computational complexity.
It is also possible to combine them in the same calculation, as will be discussed in Section \ref{sec:optimisation}.

The key feature that makes it possible to split the calculation into a sequence of few-qubit calculations is the  causal cone structure of the VILMA maps represented in Fig.~\ref{fig:layers}. More precisely, the past causal cone of qubit 1 (that is, the maps on which the reduced operator $\mathrm{Tr}_{2, \ldots, 6} [\Lambda(D_{\mathbf{m}_i})]$ depends) only involves maps $\Lambda_{12}^1$, $\Lambda_{23}^1$, and $\Lambda_{12}^2$, and hence qubits 1, 2, and 3.
Likewise, the past causal cone of qubit 2 additionally involves maps $\Lambda_{34}^1$ and $\Lambda_{23}^2$, that is, it requires the inclusion of qubit 4. However, since we can previously trace out qubit 1 in the calculation, we only need to take into account non-trivial operators in the joint Hilbert space of qubits 2, 3, and 4 in this step.

In general, other circuit structures may be used for the maps in the algorithm, but the efficiency of the algorithm highly depends on the causal structure of the circuit.
For the structure considered above, the scaling of the method is linear in the number of qubits $N$.
However, as we add more layers to the map circuit (that is, sequences of maps $\Lambda_{ij}^3$, $\Lambda_{ij}^4$, etc), the past causal cone of each qubit involves more qubits (e.g., four qubits for three layers, and so on), which results in higher-dimensional intermediate operators and thus higher computational cost.
Moreover, recall that the sequential computation presented above must be performed for all POVM outcomes $\mathbf{m}_i$ and terms $\mathbf{k}$ in $\mathcal{O}$.

\section{Numerical simulations}\label{sec:numerics}
Let us illustrate the method with the following numerical experiment.
As a reference state, that is, the one in the QPU, we consider a perturbed version of the ground state of the Hamiltonian $H$ of the hydrogen molecule $\rm{H}_2$ in the 6-31G basis with a stretched geometry (bond distance $d = 2$\r{A}) mapped to qubit space using the Jordan-Wigner transformation.
The resulting Hamiltonian can be encoded with eight qubits.
As a perturbation, we consider the image $\rho$ of the ground state under a sequence of channels $\mathcal{N}_{ij}^1$ forming a circuit like that in Fig.~\ref{fig:maps}A).
In this way, by construction, there exists a single-layer VILMA circuit capable of cancelling the perturbation.
To control the magnitude of the perturbation, the $\mathcal{N}_{ij}^1$ are chosen as $\mathcal{N}_{ij}^1 = (1 - p) \mathbb{1} + p \mathcal{E}_{ij}^1$, where $\mathcal{E}_{ij}^1$ is randomly sampled from the space of two-qubit CPTP maps, and $p$ is a parameter that we fix to $p = 0.05$.

Next, we sample from $\rho$ using a POVM composed of single-qubit symmetric IC-POVM.
We sample 50 realisations of $S = 10^4$ samples each.
Our aim is to show that, for each such batch of $S = 10^4$ samples, we can compute multiple expectation values, for different operators and maps.
In particular, we consider three maps: identity, as a reference with the original perturbed state, the inverse of the perturbation, to illustrate that non-physical operations can be used and, finally, a layer of randomly chosen unitaries, which could represent a sequence of gates that one may consider applying on the quantum computer.
As operators, we consider the Hamiltonian, along with the real parts of  1-body and  2-body reduced density matrices, $\mathrm{Re} \langle a_i^\dagger a_j \rangle$ and $\mathrm{Re} \langle a_i^\dagger a_j a_k^\dagger a_l \rangle$.

\begin{figure}[t]
    \centering
    \includegraphics[width=\columnwidth]{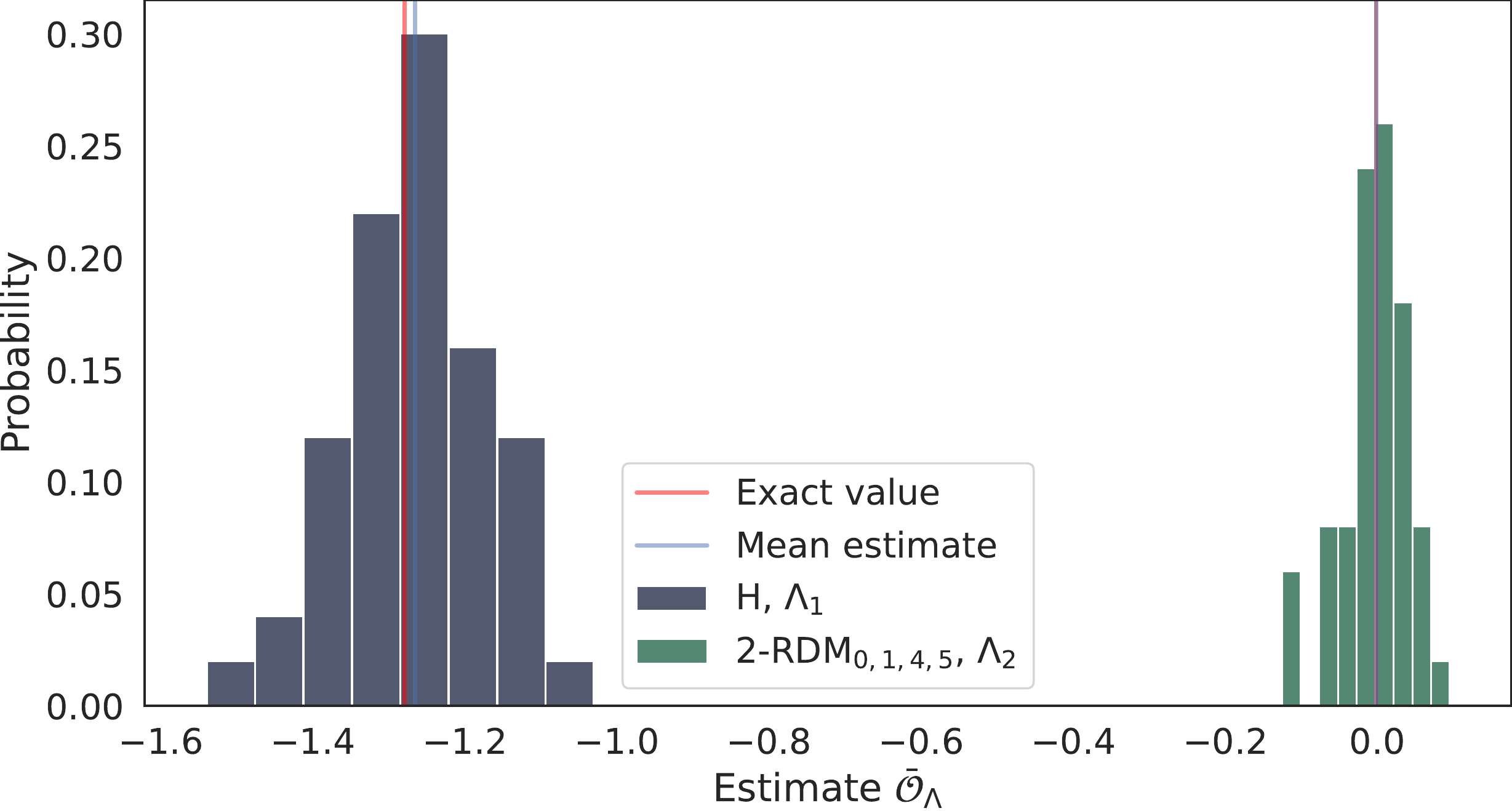}
    \caption{Statistical estimations of three observables, the Hamiltonian $H$, a 1-RDM element $\mathrm{Re} \langle a_2^\dagger a_7 \rangle$, and a 2-RDM element $\mathrm{Re} \langle a_0^\dagger a_1 a_4^\dagger a_5 \rangle$, using three maps, the identity map $\Lambda_0$, the inverse of the perturbation map $\Lambda_1$, and a layer of random unitaries $\Lambda_2$, over 50 realisations of $S = 10^{4}$ measurement rounds each.
    The histograms show the distribution of estimates $\bar{\mathcal{O}}_\Lambda$ for two different pairs of observable and map, along with the exact value (red vertical line).
    The blue vertical lines indicate the mean of the estimations over the 50 realisations.
    }
    \label{fig:histograms}
\end{figure}

\begin{figure}[t]
    \centering
    \includegraphics[width=\columnwidth]{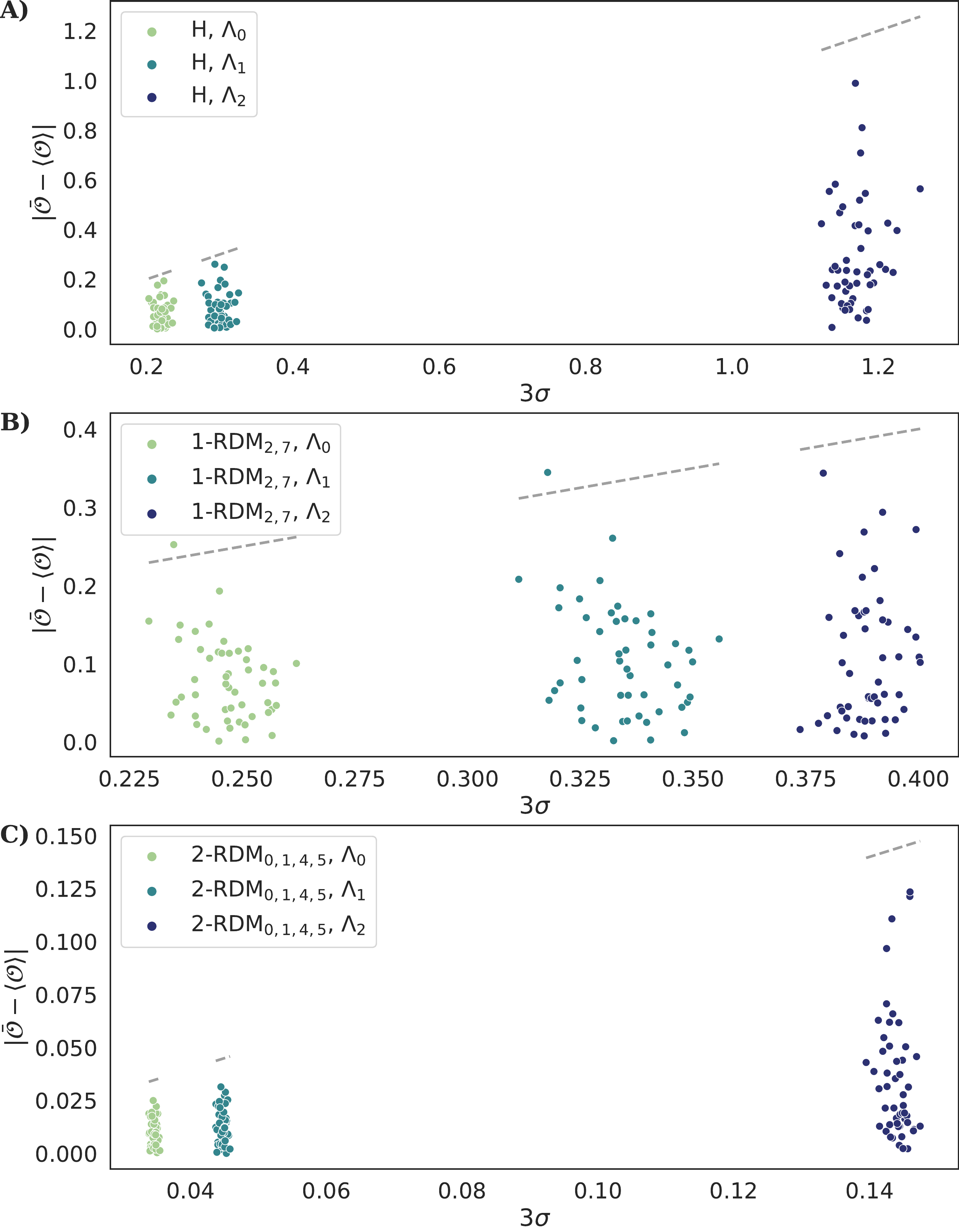}
    \caption{Statistical estimations of three observables, the Hamiltonian $H$, a 1-RDM element $\mathrm{Re} \langle a_2^\dagger a_7 \rangle$, and a 2-RDM element $\mathrm{Re} \langle a_0^\dagger a_1 a_4^\dagger a_5 \rangle$, using three maps, the identity map $\Lambda_0$, the inverse of the perturbation map $\Lambda_1$, and a layer of random unitaries $\Lambda_2$, over 50 realisations of $S = 10^{4}$ measurement rounds each.
    A--C) With the sampling data obtained in every realisation, we compute $\bar{\mathcal{O}}_\Lambda$ for every combination of map and observable and, with it, the error in the estimation, $\vert \bar{\mathcal{O}}_\Lambda - \mathcal{O}_\Lambda \vert$.
    We also compute each estimate of the error $\sigma$.
    With high probability, the error $\vert \bar{\mathcal{O}}_\Lambda - \mathcal{O}_\Lambda \vert$ should be smaller than $3 \sigma$.
    To show that this is indeed the case, for every realisation, and for every pair $(\mathcal{O}, \Lambda)$, we draw a point $(3 \sigma, \vert \bar{\mathcal{O}}_\Lambda - \mathcal{O}_\Lambda \vert)$.
    It can be seen that in nearly all cases, points lie below the diagonal (dashed line).
    }
    \label{fig:3sigmas}
\end{figure}

In Fig.~\ref{fig:histograms}, we show the resulting distribution of $\bar{\mathcal{O}}_{\Lambda}$ for the different maps and observables.
The red and blue vertical lines indicate the exact values $\mathcal{O}_{\Lambda}$ and the average of $\bar{\mathcal{O}}_{\Lambda}$ over realisations, respectively. 
The obtained values fluctuate around the correct value.
Importantly, as discussed in Section \ref{sec:method}, the method also produces, along with the estimate $\bar{\mathcal{O}}_{\Lambda}$, an estimation of the error incurred, $\sigma \equiv \sqrt{\bar{V}(\mathcal{O}_{\Lambda}) / S}$.
To assess whether the error estimations $\sigma$ are meaningful, we compare, for each estimation, the actual error and the estimated one.
In particular, the error $|\bar{\mathcal{O}}_{\Lambda} - \mathcal{O}_{\Lambda}|$ should be smaller than $3 \sigma$ with high probability.
In Fig.~\ref{fig:3sigmas} we show that, indeed, for nearly all estimations, this is the case.

Importantly, while VILMA allows us to reuse the IC measurement data to estimate many expectation values on many different states, these estimations are statistically correlated.
Thus, one should keep in mind that $\sqrt{\bar{V}(\mathcal{O}_{\Lambda}) / S}$ yields an estimate of the error of $\bar{\mathcal{O}}_{\Lambda}$, but two different estimations $\bar{\mathcal{O}}_{\Lambda}$ and $\bar{\mathcal{O'}}_{\Lambda'}$ will generally have non-zero covariance.

\section{Variational optimisation with VILMA}\label{sec:optimisation}
For some applications, VILMA may be used as a classical boost to a variational calculation.
In such situations, we are interested in finding a map $\Lambda$ that minimises or maximises an observable average $\mathcal{O}_\Lambda$ for the given input state.
In order to carry out the optimisation efficiently, we can also use the specific structure of the VILMA circuit to evaluate the observable average as a function of a single map component $\Lambda_{ij}^l$, while keeping all other components fixed.
It is therefore unnecessary to repeat the whole algorithm presented above for all the other maps at each optimisation step if only $\Lambda_{ij}^l$ is modified.
In fact, it is possible to write down an expression, linear in $\Lambda_{ij}^l$, for $\bar{\mathcal{O}}_\Lambda$ that only depends on operators defined in the local Hilbert space of qubits $i$ and $j$.
In what follows, we outline how this can be done.

Suppose we want to optimise an observable average with respect to a specific map component $\Lambda_{ij}^l$.
As discussed above, VILMA can be applied both in the forwards and backwards directions.
If both directions can be applied efficiently, as for the symmetric structure discussed in the previous section, the strategy is to apply the algorithm forwards until the stage in which $\Lambda_{ij}^l$ would be applied, and then backwards for the rest of the maps, so that eventually we arrive at an expression of the form $\mathrm{Tr} [\Lambda(D_{\mathbf{m}_i}) P_{\mathbf{k}}] = \sum_{a} \mathrm{Tr} [\Lambda_{ij}^l(R_a^{(\mathbf{m}_i, \mathbf{k})}) \bar{R}_a^{(\mathbf{m}_i, \mathbf{k})}]$.
The estimator as a function of $\Lambda_{ij}^l$ then reads
\begin{equation}\label{eq:local_opt}
    \bar{\mathcal{O}}_\Lambda = \sum_{i = 1}^S \frac{1}{S} \sum_{\mathbf{k}} c_\mathbf{k} \sum_{a} \mathrm{Tr} [\Lambda_{ij}^l(R_a^{(\mathbf{m}_i, \mathbf{k})}) \bar{R}_a^{(\mathbf{m}_i, \mathbf{k})}]
\end{equation}
The linearity of this expression enables using linear programming, including semidefinite programming (SDP) if, for instance, one is interested in imposing the complete positivity of the map $\Lambda_{ij}^l$.
We illustrate the derivation of the above expression with an example in what follows, but the generalisation to other situations should be clear.

Consider again the map in Fig.~\ref{fig:layers} and let us single out the map $\Lambda_{2, 3}^{2}$.
In the calculation of $\mathrm{Tr} [\Lambda(D_{\mathbf{m}_i}) P_{\mathbf{k}}]$, the forward algorithm starts by computing $R_{2, 3}$ (Eq.~\eqref{eq:R23}).
Next, we would need to apply $\Lambda_{34}^1$ and $\Lambda_{23}^2$ to $R_{2, 3} \otimes D_{m_4}$.
Instead, however, we apply only $\Lambda_{34}^1$ and calculate explicitly
\begin{equation}
    R_{2,3,4} = \mathbb{1}_{q_2} \otimes \Lambda_{34}^1 (R_{2, 3} \otimes D_{m_4}).
\end{equation}
The backwards algorithm is then applied until qubit 5 is traced out, leaving the residual operator $\bar{R}_{3, 4}$.
Let us define
\begin{equation}
    \bar{R}_{2,3,4} = P_{k_2} \otimes \bar{R}_{3, 4}.
\end{equation}
Notice that we can now write
\begin{equation}
    \mathrm{Tr} [\Lambda(D_{\mathbf{m}_i}) P_{\mathbf{k}}] = \mathrm{Tr} [\Lambda_{23}^2 \otimes \mathbb{1}_{q_4} (R_{2,3,4}) \bar{R}_{2,3,4} ].
\end{equation}
Finally, if we write $R_{2, 3, 4} = \sum_a R_a^{(\mathbf{m}_i, \mathbf{k})} \otimes B_a$ and $\bar{R}_{2, 3, 4} = \sum_a \bar{R}_a^{(\mathbf{m}_i, \mathbf{k})} \otimes B_a$, where $\{ B_a \}$ is the normalised Pauli basis in the Hilbert space of qubit 4, on which $\Lambda_{23}^2$ acts trivially, we see that $\mathrm{Tr} [ \Lambda_{23}^2 \otimes \mathbb{1}_{q_4} (R_{2,3,4}) \bar{R}_{2,3,4} ] = \sum_{a, a'} \mathrm{Tr} [\Lambda_{23}^2(R_a^{(\mathbf{m}_i, \mathbf{k})}) \bar{R}_{a'}^{(\mathbf{m}_i, \mathbf{k})} \otimes B_a B_{a'} ] = \sum_{a} \mathrm{Tr} [\Lambda_{23}^2(R_a^{(\mathbf{m}_i, \mathbf{k})}) \bar{R}_a^{(\mathbf{m}_i, \mathbf{k})}]$.
This procedure allows us to pre-compute a set of low-dimensional operators, $\{ R_a^{(\mathbf{m}_i, \mathbf{k})}, \bar{R}_a^{(\mathbf{m}_i, \mathbf{k})} \}_{a, (\mathbf{m}_i, \mathbf{k})}$, that capture 
all the dependence of $\bar{\mathcal{O}}_\Lambda$ on a specific map component $\Lambda_{ij}^l$.

The algorithm above can be summarised in the following general steps:
1) run the forward algorithm, stopping right before the singled-out map $\Lambda_s$ is applied. Decompose the resulting operator $R$ as $R = \sum_a R_a^{(\mathbf{m}_i, \mathbf{k})} \otimes B_a$, where $\{ B_a \}$ is the normalised Pauli basis in the Hilbert space of the qubits on which $\Lambda_s$ acts trivially, and store $\{ R_a^{(\mathbf{m}_i, \mathbf{k})} \}$.
2) run the backward algorithm, stopping right before the singled-out map $\Lambda_s$ is applied. Decompose similarly the resulting operator $\bar{R}$ as $\bar{R} = \sum_a \bar{R}_a^{(\mathbf{m}_i, \mathbf{k})} \otimes B_a$ and store $\{ \bar{R}_a^{(\mathbf{m}_i, \mathbf{k})} \}$.
Repeating the process for every measurement outcome and Pauli string pair $(\mathbf{m}_i, \mathbf{k})$ allows us to use Eq.~\eqref{eq:local_opt} for optimisation.

\begin{figure}[t!]
    \centering
    \includegraphics[width=\columnwidth]{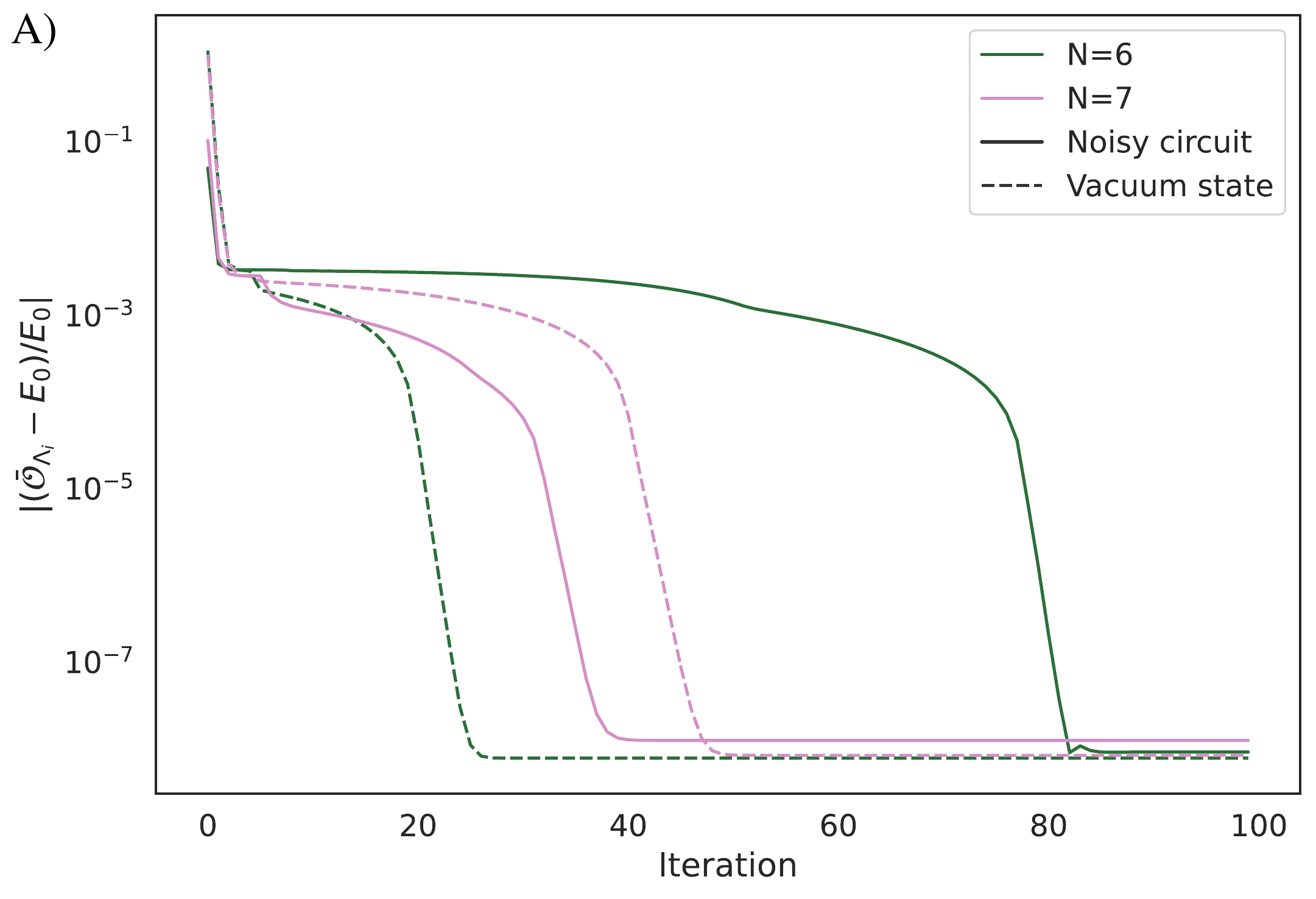}
    \includegraphics[width=\columnwidth]{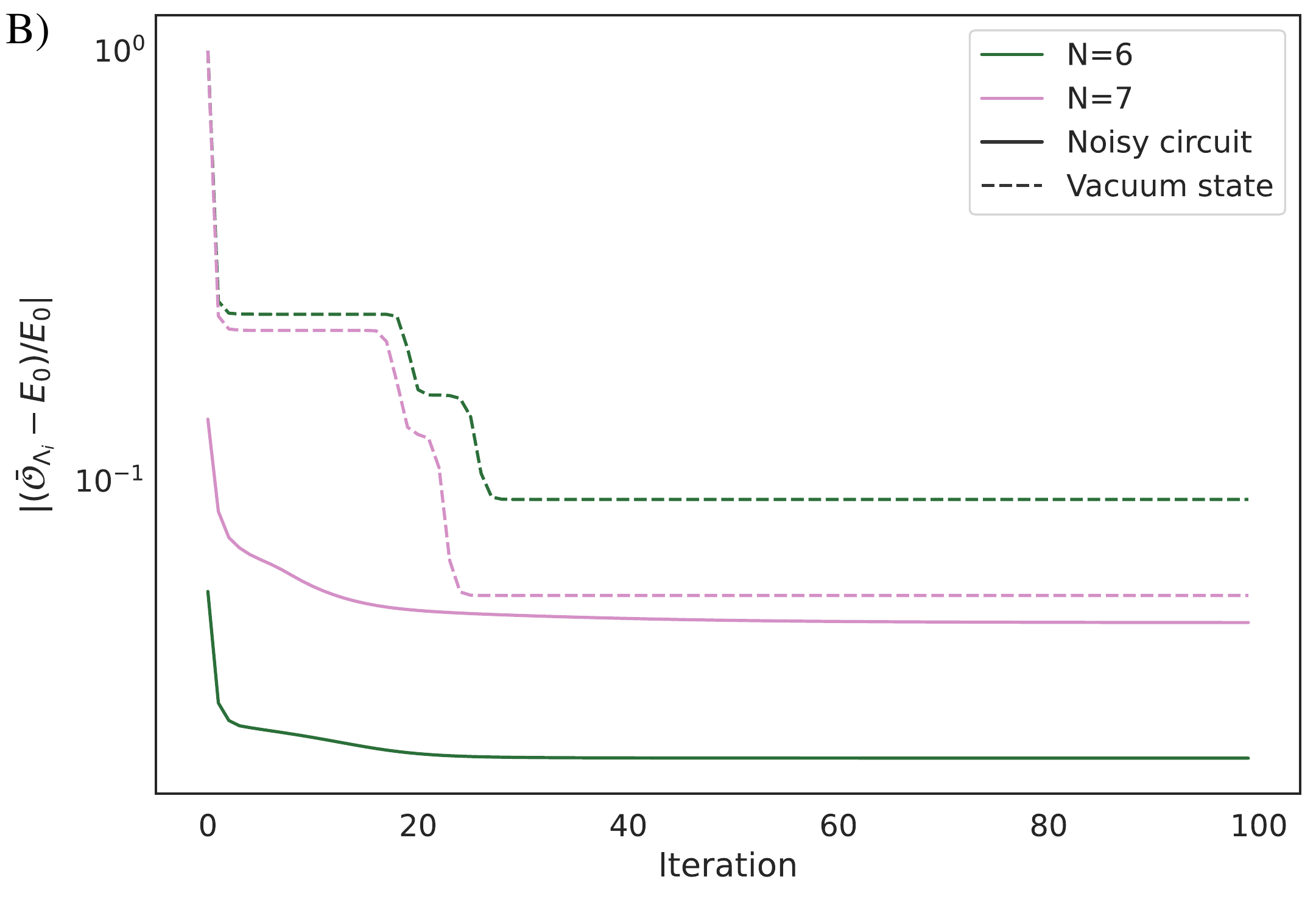}
    \caption{Relative error between the energy of the image of the input state through the VILMA map, $\bar{\mathcal{O}}_{\Lambda}$, and the ground state energy, $E_0$, along the optimisation process for two input states, a noisy VQE simulation and $\ket{0}^{\otimes N}$, and four XX model Hamiltonians, for two system sizes and two values of the magnetic field.
    A) Right below the phase transition of the model, the energy converges to within the numerical precision of the optimiser, regardless of the input state.
    The number of iterations required changes notably from one example to another.
    For $N=6$, convergence is reached faster with the input vacuum state than with the noisy VQE input, while the opposite is true for $N=7$.
    B) For the zero-field XX model, the optimisation does not reach the ground state energy.
    However, with the noisy circuit input state, VILMA decreases the energy with respect to the initial reference state and, in both cases, below the energy that can be reached using the classical input state $\ket{0}^{\otimes N}$.
    }
    \label{fig:variational_vilma}
\end{figure}

We apply this procedure on a single-layer VILMA circuit for four different XX model Hamiltonians, $H = - J [\sum_i (\sigma_x^{(i)} \sigma_x^{(i+1)} + \sigma_y^{(i)} \sigma_y^{(i+1)}) / 2 + B \sigma_z^{(i)}]$, where $J$ is the coupling constant, that we set to one, $B$ is the magnetic field, and $\sigma_k^{(i)}$ with $k = x, y, z$ are Pauli matrices.
Periodic boundary conditions, $\sigma_k^{(N + 1)} = \sigma_k^{(1)}$, are assumed.
The cases that we consider correspond to $B=0$ and $B \approx 1$ (right below the phase transition exhibited by the model in the thermodynamic limit~\cite{son_quantum_2009}).
At finite size, the ground state exhibits a series of level crossings as a function of $B$, as a result of which the entanglement structure of the state changes notoriously~\cite{son_quantum_2009,sokolov_emergent_2022}.
We also use two different system sizes, $N = 6$ and $N = 7$ spins.

As reference states we take the outputs of noisy circuits.
The circuits are pre-trained VQE ans\"{a}tze that, in the absence of noise, would  approximate the ground states~\cite{bharti_noisy_2022}.
However, we apply noisy gates.
Every CNOT is composed with a channel $\mathcal{N}_{\mathrm{CNOT}}$, which itself is a composition of depolarising noise (with $p = 10^{-3}$) and coherent noise (the two qubits are rotated by applying two single-qubit rotations $R_x(\theta) R_z(\theta)$ with an angle $\theta = 0.05$ to each of them.

The single-layer VILMA map is optimised in the following way.
Every map $\Lambda_{ij}^1$ is chosen sequentially and, for each of them, we compute the operators $\{ R_a^{(\mathbf{m}_i, \mathbf{k})}, \bar{R}_a^{(\mathbf{m}_i, \mathbf{k})} \}_{a, (\mathbf{m}_i, \mathbf{k})}$ to then minimise $\bar{\mathcal{O}}_{\Lambda}$ over $\Lambda_{ij}^1$.
In the optimisation, we impose each map 
to be CPTP to ensure that the output operator is a valid state.
The resulting constrained optimisation of Eq.~\eqref{eq:local_opt} is an SDP, which can be solved efficiently.
Since we are now only interested in illustrating the local optimisation strategy described earlier, in this artificial numerical experiment we choose to bypass the additional complications stemming from the optimisation with finite statistics (\eg over-fitting) by considering the exact noisy state as input, that is, $\rho$ instead of $\rho_S$.
In practice, however, finite statistics-related issues need to be handled.

In Fig.~\ref{fig:variational_vilma}, we plot the relative error of the resulting step at each iteration (solid lines).
For $B \approx 1$, the optimisation finds the ground state within the numerical precision of the optimiser for both system sizes.
For $B = 0$, on the other hand, it finds lower energies, but does not reach the ground state. Importantly, recall that the optimisation is constrained to the complete positivity of $\Lambda$ because we are neglecting all information about the origin of the input state and still enforcing the positivity of the output one.
As a consequence, even if a single-layer map mapping the input state to the ground state exists, it may not be CP, and hence remain inaccessible to the optimisation procedure.
In practice, one should use strategies in which information about the noise in the device is taken into account, which is in principle possible given that VILMA per se does not require the positivity of the maps.
In any case, notice that even with one layer of maps and modest classical compute (indeed, we never reconstruct operators of dimension larger than $4 \times 4$), the algorithm could boost the result obtained with a quantum computer.
Once again, we stress that, in real experiments, this will require managing additional finite statistics-related matters.

\section{VILMA as a classical ansatz}\label{sec:classicalVILMA}

So far, we considered VILMA as a classical algorithm that takes the measurement statistics from a quantum processor as an input.
Another possibility would be to input the classical description of a quantum state (for instance $\ket{0}^{\otimes N}$) and look for the optimal VILMA maps that, say, produce a state achieving the minimum expected value of a given observable (\eg a Hamiltonian).
In this way, VILMA can be used as an algorithm to produce a classically efficient ansatz.

In Fig.~\ref{fig:variational_vilma}, we compare this technique with the results obtained in the previous section.
On the one hand, we can see that, for $B$ nearing the transition, VILMA can find the ground state even with the $\ket{0}^{\otimes N}$ input state.
However, if the maps are initially set to identity, the local optimisation sweeps are futile, as the system seems to be at a local minimum.
By initialising the VILMA map with randomly chosen unitary gates, the problem is overcome. This indicates that there is potential room for improvement in the optimisation strategy. In fact, finding the ground state with a single-layer of VILMA should be expected, given that the ground state of the Hamiltonian is a W state, which can be prepared by a single layer of unitary gates~\cite{cruz_efficient_2019}.

We also notice that in Ref.~\cite{Ran_2020} it is shown that, in the case of a layered circuit of two-qubit maps like the one employed here set to be unitary operations, the produced state can encode matrix product states \cite{Cirac_2021} whose bond dimension depends on the number of layers in the circuit.
The purely classical version of VILMA can go beyond this in two ways: (i) it is not constrained to unitary maps, and (ii) different topologies can be considered.
We leave as an open question the precise characterisation of the class of states reachable with classical VILMA and the connection with MPS and other tensor network methods.

In any case, it is worth discussing the connection between VILMA with classical and quantum inputs.
In particular, we make the following observation: if a state $\rho'$ is classically reachable with VILMA, that is, if we can access it by applying an affordable VILMA map $\Lambda$ to $\ket{0}^{\otimes N}$, then $\rho'$ is reachable with the same VILMA map structure if the input is the state of the quantum processor $\rho$.
This is so because it is possible to compose the maps in the first layer of VILMA with single-qubit CPTP channels that map any input state to $\ket{0}$ in such a way that the $N$-qubit input state is initially mapped to $\ket{0}^{\otimes N}$~\footnote{For instance, for a layered structure like the one in Fig.~\ref{fig:layers}, this can be done in the following way.
Let $\Lambda$ be a map such that $\rho' = \Lambda(\ketbra{0}^{\otimes N})$, and $\mathcal{Z}$ be the single-qubit CPTP map defined by $\mathcal{Z}(\sigma) = \ketbra{0}$.
Now, consider the map $\Lambda'$ with the same structure as $\Lambda$ and in which $\Lambda_{ij}^{'l} = \Lambda_{ij}^{l}$ for $l > 1$.
For the first layer $l = 1$, let $\Lambda_{12}^{'1} = \Lambda_{12}^{1} \circ (\mathcal{Z}_{1} \otimes \mathcal{Z}_{2})$ and $\Lambda_{ij}^{'1} = \Lambda_{ij}^{1} \circ (\mathbb{1}_{i} \otimes \mathcal{Z}_{j})$ for all other maps in the layer.
The map $\Lambda'$ maps any input state to $\rho'$.}.
The converse is clearly not true, 
since the input data itself can correspond to a state not classically reachable by VILMA.
Therefore, VILMA may enable computing observable averages that may be inaccessible using only classical methods or only limited quantum resources.
The results in Fig.~\ref{fig:variational_vilma} B) can be interpreted as an example of this: the algorithm can benefit from the correlations in the input state to reach states (in this case, lower-energy ones) that would not be accessible with the same structure using a product input state.
In particular, in the simulation for $N = 7$, the state resulting from the noisy circuit execution has a higher energy than what can be achieved with a single-layer VILMA map.
While one would deem such quantum computer output useless, given that a better result is achievable classically, the algorithm exploits the data to achieve a lower energy than with the separable input.

\section{Conclusions} \label{sec:conclusion}

In this paper we have introduced VILMA, a scalable algorithm to efficiently apply a sequence of (not necessarily physical) linear maps on quantum states and estimate the result of measurements applied on this modified state. Most importantly, VILMA can be used in the case where the states under scrutiny are not known, but we have access to just a finite sample of outcomes of a tomographically complete set of measurements applied to these states. Yet, the resulting estimates of expectation values are unbiased, and the algorithm provides a meaningful estimate of the error incurred due to finite statistics. This allows to use VILMA to post-process the results of a quantum processor in a regime where performing full state quantum tomography is out of reach.

Besides estimating mean values of post-processed states, VILMA also 
enables optimising over the sequence of linear maps, opening the possibility for applying it to several ends, such as noise mitigation and variational search of Hamiltonian ground states. Our results suggest that, on the one hand, this method has a very natural classical counterpart, in which the input state is a product state, that may have interesting connections with matrix product states or other tensor network methods.
We have shown that purely classical VILMA can indeed find ground states of many-body Hamiltonians efficiently in some cases.
On the other hand, the algorithm can take as input the IC measurement data from a quantum computer and boost the results, and we have discussed that, setting optimisation-related issues aside, the class of states reachable with such inputs is strictly larger than using product input states.
We have provided proof-of-concept simulation results in which noisy circuit states are used to achieve lower energies than can be found with product input states.

There are several open questions related to VILMA that we leave for future investigations. First of all, one must be careful when using VILMA variationally on applications involving real experimental data to not over-fit the results due to statistical fluctuations. One possibility to avoid this is to perform the optimisation of the maps on a different set of experimental data than the ones used for the final estimation of the expected values of interest. Another interesting line of research is to study to what extent VILMA can be used as a noise mitigation scheme for realistic types and levels of noise in current hardware. Finally, the connection between VILMA, MPS and other ans\"{a}tze is certainly worth further investigation. 

\textbf{Acknowledgements:} We acknowledge Zolt\'{a}n Zimbor\'{a}s, Adam Glos, and Stefan Knecht for interesting discussions.
\textbf{Competing interests:} Elements of this work are included in a patent filed by Algorithm Ltd with the European Patent Office.
\textbf{Authors contributions:} GGP conceived the algorithm.
EMB, DC, GGP designed and directed the research.
GGP, ML, JM, MACR, BS implemented the algorithm and ran simulations.
EMB, DC, GGP wrote the first version of the manuscript.
All authors contributed to scientific discussions and to the writing of the manuscript.

\appendix

\section{Further mathematical details}\label{app:formalisation}

The main text intended to introduce the method to calculate the traces in an intuitive manner with an explicit example.
In this section, we formalise the ideas in more detail.

Let $\Lambda$ be a linear map acting in the space of linear operators $L( \bigotimes_{i=1}^{N} \mathcal{H}_i)$ of a set of $N$ quantum systems $\lbrace q_1, \ldots, q_N \rbrace$.
The map is composed of $L$ $k$-local maps, $\Lambda = \bigcirc_{r = 1}^{L} \Lambda_r \equiv \Lambda_{L} \circ \cdots \circ \Lambda_{2} \circ \Lambda_{1}$, that is, $\Lambda_r$ acts non-trivially on at most $k$ subsystems.
Since some of the map components may commute with one another, the map $\Lambda$ may admit multiple such decompositions.
In particular, let $\mathcal{T} = \lbrace T \in S_L | \bigcirc_{r = 1}^{L} \Lambda_{T(r)} = \Lambda \rbrace$, where $S_L$ is the permutation group of $L$ elements, be the set of all possible permutations of map components that preserve $\Lambda$.

Consider an element $T \in \mathcal{T}$, and an integer $1 \leq l < L$ such that $\bigcirc_{r = l + 1}^{L} \Lambda_{T(r)}$ acts trivially on some subsystem $q_i$. Denote $\Lambda^{\bar{q}_i}
\equiv \bigcirc_{r = l + 1}^{L} \Lambda_{T(r)}$ and $\Lambda^{q_i}
\equiv \bigcirc_{r = 1}^{l} \Lambda_{T(r)}$.
Obviously, we have $\Lambda = \Lambda^{\bar{q}_i} \circ \Lambda^{q_i}$.
Furthermore, let $\mathcal{F}(\Lambda^{q_i}) \subset \{ 1, \ldots, N \}$ indicate the set of subsystems on which $\Lambda^{q_i}$ acts non-trivially and $\bar{\mathcal{F}}(\Lambda^{q_i}) = \{ 1, \ldots, N \} \setminus \mathcal{F}(\Lambda^{q_i})$ its complement.

To simplify the notation, we now denote the dual effects by $D = \bigotimes_{i=1}^{N} D_i$ and the Pauli strings by $P = \bigotimes_{i=1}^N P_i$.
We can write
\begin{widetext}
\begin{equation}
    \begin{aligned}
    \mathrm{Tr}[\Lambda(D) P] &= \mathrm{Tr} \left[ \Lambda^{\bar{q}_i} \left( \Lambda^{q_i} \left( \bigotimes \limits_{k \in \mathcal{F}(\Lambda^{q_i})} D_k \right) \otimes \bigotimes \limits_{k' \in \bar{\mathcal{F}}(\Lambda^{q_i})} D_{k'} \right) \bigotimes \limits_{k'' = 1}^{N} P_{k''} \right] \\
    \end{aligned}
\end{equation}
In order to make the next steps clear and explicit, it is useful to write, using the fact that $\Lambda^{\bar{q}_i}$ acts trivially on subsystem $q_i$, $\Lambda^{\bar{q}_i} (\cdot) = \sum_{a, b} \lambda_{ab} \mathbb{I}_i \otimes B_a \cdot \mathbb{I}_i \otimes B_b^\dagger$, where $\{ B_a \}$ is a basis of $L(\bigotimes_{k \neq i} \mathcal{H}_k)$.
Therefore, we have
\begin{equation}
    \begin{aligned}
    \mathrm{Tr}[\Lambda(D) P] &= \sum_{a, b} \lambda_{ab}\mathrm{Tr} \Bigg[ \left( \mathbb{I}_i \otimes B_a \right) \left( \Lambda^{q_i} \left( \bigotimes \limits_{k \in \mathcal{F}(\Lambda^{q_i})} D_k \right) \otimes \bigotimes \limits_{k' \in \bar{\mathcal{F}}(\Lambda^{q_i})} D_{k'} \right) \left( \mathbb{I}_i \otimes B_b^\dagger \right)  \left( P_i \otimes \mathbb{I} \right) \left( \mathbb{I}_i \otimes \bigotimes \limits_{k'' \neq i} P_{k''} \right) \Bigg] \\
    &= \sum_{a, b} \lambda_{ab}\mathrm{Tr} \left[ B_a \left(  \mathrm{Tr}_i \left[ \left( P_i \otimes \mathbb{I} \right) \Lambda^{q_i} \left( \bigotimes \limits_{k \in \mathcal{F}(\Lambda^{q_i})} D_k \right) \right] \otimes \bigotimes \limits_{k' \in \bar{\mathcal{F}}(\Lambda^{q_i})} D_{k'} \right) B_b^\dagger \bigotimes \limits_{k'' \neq i} P_{k''} \right] \\
    \end{aligned}
\end{equation}
\end{widetext}
Let us now redefine $\Lambda^{\bar{q}_i} (\cdot) = \sum_{a, b} \lambda_{ab} B_a \cdot B_b^\dagger$ and denote $R_{\mathcal{F}(\Lambda^{q_i}) \setminus \{ i \}} = \mathrm{Tr}_i \left[ \left( P_i \otimes \mathbb{I} \right) \Lambda^{q_i} \left( \bigotimes \limits_{k \in \mathcal{F}(\Lambda^{q_i})} D_k \right) \right]$.
The above expression now reads
\begin{equation}
    \mathrm{Tr}[\Lambda(D) P] = \mathrm{Tr} \left[ \Lambda^{\bar{q}_i} \left( R_{\mathcal{F}(\Lambda^{q_i}) \setminus \{ i \}} \otimes \bigotimes \limits_{k \in \bar{\mathcal{F}}(\Lambda^{q_i})} D_{k} \right) \bigotimes \limits_{k' \neq i} P_{k'} \right].
\end{equation}

Writing $\Lambda^{\bar{q}_i} = \Lambda^{\bar{q}_j} \circ \Lambda^{q_j}$ where, this time, $\Lambda^{\bar{q}_j}$ acts trivially on subsystem $q_j$, we can iterate the process by applying $\Lambda^{q_j}$ to $R_{\mathcal{F}(\Lambda^{q_i}) \setminus \{ i \}} \otimes \bigotimes_{k \in \bar{\mathcal{F}}(\Lambda^{q_i})} D_{k}$ (or, rather, to the relevant subsystems $( \mathcal{F}(\Lambda^{q_i}) \setminus \{ i \} ) \cup \mathcal{F}(\Lambda^{q_j})$ if $( \mathcal{F}(\Lambda^{q_i}) \setminus \{ i \} ) \cap \mathcal{F}(\Lambda^{q_j}) \neq \emptyset$ or to $\mathcal{F}(\Lambda^{q_j})$ otherwise) and taking the corresponding partial trace over $q_j$.

Importantly, the highest-dimensional operators that must be computed in the above calculation are the terms such as $\left( P_i \otimes \mathbb{I} \right) \Lambda^{q_i} \left( \bigotimes_{k \in \mathcal{F}(\Lambda^{q_i})} D_k \right)$.
Thus, the efficiency of the algorithm relies on the appropriate choice of integer $l$ at each step (typically, a value that minimises the number of subsystems on which the corresponding $\Lambda^{q_i}$ acts, that is, $\vert \mathcal{F}(\Lambda^{q_i}) \vert$, is desirable).
Similarly, in order to make sure that the procedure only involves the explicit calculation of operators of small enough dimension, a proper choice of map structure and permutation $T \in \mathcal{T}$ are necessary as well.
Indeed, notice that the residual operators $R$ live in the Hilbert space of the subsystems in the union of intermediate relevant spaces (minus the traced-out parties).
Hence, it is in general a good idea to design the map structure so that only one party needs to be added in every iteration, so the partial trace balances the total count.
The text presents an example in which all the elements have been chosen in such a way that the computation is efficient for any number of qubits $N$.



%

\end{document}